\documentclass[aps,prd,twocolumn,groupedaddress,eqsecnum,amssymb,showpacs,
nofootinbib]{revtex4}
\usepackage{graphicx,mathptm,bm,times,epsfig}

\newcommand{\bea}{\begin{eqnarray}}
\newcommand{\eea}{\end{eqnarray}}

\begin{document}

\title{Notes on Time's Enigma}

\author{Laura Mersini-Houghton}
\email[]{mersini@physics.unc.edu}
\affiliation{Department of Physics and Astronomy, UNC-Chapel Hill,
NC, 27599-3255, USA \\and,\\ Department of Applied Mathematics and Theoretical Physics, Cambridge University, Cambridge, U.K.}

\date{\today}
 
\begin{abstract}

Scientists continue to wrestle with the enigma of time. Is time a dynamic or a fundamental property of spacetime? Why does it have an arrow pointing from past to future? Why are physical laws time-symmetric in a universe with broken time-reversal symmetry? These questions remain a mystery. The hope has been that an understanding of the selection of the initial state for our universe would solve such puzzles, especially that of time's arrow. 

In this article, I discuss how the birth of the universe from the multiverse helps to unravel the nature of time and the reasons behind the time-reversal symmetry of our physical laws. I make the distinction between a local emerging arrow of time in the nucleating universe and the fundamental time with no arrow in the multiverse. The very event of nucleation of the universe from the multiverse breaks time-reversal symmetry, inducing a locally emergent arrow. But, the laws of physics imprinted on this bubble are not processed at birth. Time-reversal symmetry of laws in our universe is inherited from its birth in the multiverse, since these laws originate from the arrowless multiversal time.

\end{abstract}


\maketitle

\section{Introduction} 
\label{sec:intro}

Time - the enigmatic building block of the cosmos - has stubbornly challenged natural philosophers and scientists over millenia. What is time? Why does it have an arrow? Why isn't time's arrow 'DNA-ed' into our physical theories? Such basic questions that touch upon one of nature's most fundamental properties remain mysterious.

The complexity of time's mystery becomes more enticing within the multiverse framework.
I have been advocating the necessity of viewing the cosmos as a multiverse since the advent of the landscape of string theory. The reason is: an investigation of why we started with this universe \cite{laurarich} necessarily leads  to the question, 'as compared to what other possible universes?'\cite{lauramulti}. The investigation of the birth of our universe from the landscape multiverse studied in \cite{laurarich,lauramulti,tomo}, and described briefly in the next section, shows that the selection of the initial states for universes born from the multiverse is governed by the dynamics of matter and gravitational degrees of freedom (D.o.F) and their entanglement with the background multiverse. Their birth is neither a special event nor is it occuring at a special moment. Nonequlibrium dynamics of these initial states leads to a superselection rule that picks only the high energy states as 'survivor' universes. Since the progress with the puzzle of the selection of the initial state of the universe \cite{lauramulti,laurarich,tomo} and time's enigma are intertwined, then an extension of physics into the  multiverse framework allows for deeper insights into a conceptual understanding of time. 

In what follows, the fundamental time in the multiverse is distinct from the local time in the nucleating bubble universes. This article argues that fundamental time does not have an arrow. But, that an arrow of time emerges only locally at the bubble location due to the breaking of time-reversal symmetry by the out-of-equilibrium correlations between various D.o.F's of the bubble entangled with the multiverse. Through this approach \cite{lauramulti,laurarich} an arrow of time and physical laws with time reversal symmetry can be concomitant.

\section{The Three Enigmas of Time and the Multiverse}

Time's enigma is comprised of three basic questions: A) Why do we have an arrow of time; B) What is time, fundamental or emergent; and, C) Why are physical laws time-symmetric, i.e. independent of the arrow of time?

The first question is closely related to the selection of the initial conditions of the universe. In Sec.2.A. I argue  that the arrow emerges at the moment of the bubble nucleation because the entanglement of the initial state with the multiverse and the state's nonequilibrium gravitational dynamics, create an information loss about the underlying reality. The information loss about the multiverse breaks the time-reversal symmetry at the bubble.

The second question is still open and debated. However, when the nature of time is treated within the multiverse framework, we may be in a position to draw more specific conclusions. Based on the conservation of the total information in the multiverse, the only two options left by the reversal-symmetry of this conservation are: either time is fundamental; or, it does not exist at all. I reason in Sec.2.B. that time in the multiverse is fundamental rather than nonexistent. Energy and information conservation lead to time-translation and time-reversal symmetries. That is, multiversal time becomes a fundamental building block of the cosmos. Symmetries ensure fundamental time has no direction, no beginning and no end. Fundamental time is not the same as the local time at the bubble nucleations since the latter is dynamic, breaks reversal symmetry and experiences an emergent arrow.

We now have a way of addressing the third (and probably the toughest) question, the time-reversal symmetry of the physical laws in a universe where the reversal symmetry is badly broken. As discussed in Sec.2.C., when treated in a multiverse framework, fundamental time is directionless and consequently physical laws inherit its time-reversal symmetry.  Despite that reversal symmetry is broken for the local time by the bubble nucleation, the bubble still inherits laws of physics at birth from the multiverse, without modification. Thus the emergent time's arrow in the bubble does not affect the time-reversal symmetry imprinted onto the physical laws that the bubble inherits from birth in the multiverse.
  
This article offers a way of understanding the nature of time, the emergence of its arrow and the time-reversal symmetry of physical laws in a coherent picture, by  posing time's enigma problem in the context of the multiverse.

\subsection{Time's Arrow and the Birth of the Universe}

We know what the universe looks like at present. We also experience an arrow of time from past to future. This arrow of time provides a profound insight into the initial moments of the universe. The reason is the second law of thermodynamics which leads us to conclude that time's arrow is a direct consequence of the asymmetry between the disorder of the present state and the order that must have existed in the initial state. More specifically, time's arrow implies that our universe had to start from a highly improbable state of exquisite order, with its equivalent low entropy. For this reason, the arrow is closely related to the mystery of the nucleation moment of the universe. In isolation, the second law of thermodynamics does not then resolve the enigmatic time's arrow problem but simply trades it with the enigma of what selected the initial state of the universe. But an understanding of the selection of the initial conditions of the universe would definitely represent progress in resolving the puzzle of the observed time's arrow in our universe. However, understanding time's arrow (A) is not sufficient since we still have to explore what time is (B), and why the physical laws are 'unaware' of this arrow of time (C). 

Exploring such questions requires a reconstruction of events from the present time to the Big Bang and before. As is well known, reverse-engineering is generically an ill-posed problem because a multiplicity of initial states can lead to a single present state. With this warning, even a sensible answer to time's enigma that relates it to the birth of the universe, carries a lot of ambiguity and remains in the realm of speculation until we can test the theory by experiment. 

Nevertheless, exercising caution is useful for only as long as it does not discourage scientific inquiry. With this in mind, let us start investigating time's arrow by using the progress made in \cite{laurarich} for the selection of the initial conditions of our universe from the multiverse. A knowledge of the multiverse's structure would allow us to take a top-down approach and thus bypass reverse-engineeering ambiguity. In \cite{laurarich} we used the landscape derived from string theory as our working model for the multiverse structure. For the sake of illustration, let us continue our discussion of time using the same multiverse structure, the string theory landscape. The considerations below are applicable to other types, for example eternal inflation \cite{eternal}, if their structure is known and, crucially, if the selection rule  for the surviving bubbles, (the measure), is governed by dynamics \cite{jens} instead of being fixed as an {\it apriori} initial condition. 

The question - 'why did our universe start in such a low entropy state' - was investigated and addressed in \cite{laurarich} within the framework of the landscape multiverse. I will sketch briefly the main steps and results of this program since the selection of the initial conditions mechanism is directly relevant to the study of time's arrow here.  The birth of the universe from the landscape multiverse in \cite{laurarich} was explored by proposing to place the wavefunction of the universe on the landscape multiverse, in order to study the dynamical evolution of matter and gravitational D.o.F's and their coupling to the multiverse 'bath'. The out-of-equilibrium dynamics in the initial states entangled with the multiverse 'bath' leads to a superselection rule that eliminates the possibility of low energy initial states from the phase space, and selects only the highly ordered, high energy (low entropy) states as the most probable universes. The high energy states were dubbed 'survivor universes' as they lead to the birth of physically relevant universes, and the low energy initial states were coined 'terminal universes' as they can not give rise to expanding bubbles.  The dynamics is contained in the Master Equation for the wavefunctional of the universe propagating on the multiverse. The Master Equation is a Schroedinger type equation with the gravitational and matter Hamiltonians being promoted to quantum operators. Thus it encaptures the dynamics of the wavefunctional of the universe and of the structure of the multiverse. But the Master Equation is sourced from a backreaction term of superhorizon matter modes acting on the wavefunctional. This term describes the entanglement of the multiverse 'bath' with the wavefunction, which 'pins down' the high energy branches of the wavefunctional, thereby triggering decoherence of our branch from the rest. 

Locally this initial state is a 'battlefield' that bubbles with the nonequilibrium dynamics of its matter and gravitational D.o.F's, along with the backreaction dynamics. The gravitational D.o.F.'s are captured by its vacuum energy which is trying to kick-start the initial bubble into an accelerated expansion. Entanglement with the multiverse and the backreaction of the matter D.o.F.s tries to crunch that initial state to a point. (In solid state jargon, the different behaviour of the two types of D.o.F's would be ascribed as follows: the matter D.o.F's constitute a 'positive heat capacity' system while the gravitational D.o.F's constitute a 'negative heat capacity' system. Thus the first type reaches equilbrium by driving to a crunch and the latter type reaches equlibrium by expanding to infinity. Having both types of dynamics drives the system out of equilibrium). Depending upon which one wins in this 'tug-of-war' determines whether the initial packet survives and grows to give birth to a universe or terminates in a 'stillbirth'. The high energy states can survive the backreaction of matter and the bath, and can grow to physically relevant universes. But the low energy states can not survive.  The superselection rule, derived from the nonequilibrium dynamics and entanglement with the multiverse 'bath', selects the high energy states as the 'survivor' universes and forbids the low energy 'terminal' universes. The initial phase space of all possible states for potentially starting a universe like ours, thus shrinks to the subset of high energy initial states, the 'survivor' universes. The main implication is that the phase space is not ergodic when dynamics is taken into account - an important point for the discussion of the dynamically driven asymmetry between the initial and boundary conditions below. The birth of the universe from the multiverse in this program thus offers the first explanation into the obstinate puzzle: why did our universe start in such a highly ordered (low entropy) state.  \footnote {Our program for the birth of the universe from the multiverse \cite{laurarich,lauramulti, tomo} has led to some intriguing observational consequences \cite{tomo}. Three of its predictions have already been succesfully tested so far, namely: the void \cite{rudnick}, the dark flow \cite{kashlinsky}, and $\sigma_8$ \cite{wmap,SDSS}}. This resolution for the asymmetry between the entropy of the present universe, and the reasons behind the very low entropy of the initial state, then satisfactorily addresses the observed time's arrow puzzle. 

Although this program \cite{laurarich,lauramulti,tomo} offers a natural explanation to one of the enigmas, the  arrow of time,
by facilitating our understanding of why the universe had to start in such an exquisitely ordered state of low entropy, within the multiverse framework, it is still an incomplete approach for the following reasons. The study of the dynamics of the wavefunctional of the universe in the multiverse was carried out by implictly assuming the existence of time in the multiverse. That means that we still face two further questions in relation to understanding time, namely:

i) what is time in the multiverse?
ii) why do our physical laws have a time-reversal symmetry instead of an arrow of time?

The question of what is fundamental time and the mystery of time's arrow are distinct, yet  closely related. Completing the study for the arrow of time puzzle in \cite{laurarich} by using the Master Equation of quantum mechanics to study the evolution of the initial packet, now demands that we address the issue of the existence and the nature of multiversal time. Otherwise, until a tractable understanding of the nature of time is achieved, arguments presented here and in \cite{laurarich} would become circular.

\subsection{Fundamental Time and the Multiverse}

A useful way of thinking about entropy in cosmology is as a measure of the lack of information about the underlying reality. The underlying reality here is identified with the multiverse. Information is contained in physical correlations. Correlations  are determined and quantified by physical laws. Then in principle, once the correlations are correctly identified, we should be able to estimate them.

In this discussion, energy and entropy are assumed to be meaningful concepts and, quantum mechanics is assumed to be valid should time exist. I will sometimes refer to the multiverse as the 'bath', the nucleating  universe as the 'system' and time in the multiverse as 'fundamental time'. Let us  now explore the question: 'what is time in the multiverse', i.e. does it exist, does it make sense, does it have an arrow?

By definition the multiverse is all there is. Due to the unitarity principle, the total information of the 'system' $+$ 'bath' is conserved. Since no information can be lost in the multiverse, then the only two consequent possibilites are: i) {\it either time in the multiverse does not exist} \cite{kiefer}; ii) {\it or, time in the multiverse exists as a fundamental building block of the cosmos}, with no beginning, no end and with the reversability symmetry from the conservation of information \footnote{An emerging time in the multiverse does not appear plausible since the emergence adds information on the multiverse that wasn't there prior.}. Let us now explore where the first option, namely, time does not exist, leads: if time is nonexistent in the multiverse, then all the relevant physics to us is local rather then multiversal since time evolution and dynamics, would take meaning and emerge only at the bubble nucleation. With this choice, we have no need and no means of access to the underlying reality of the multiverse since a dynamic evolution would have no prior meaning or existence. This part of nature becomes redundant and irrelevant to  a universe embedded in a {\it timeless} multiverse.

I will ascribe to the latter possibility, namely that fundamental time does exist in the multiverse because that part of reality is relevant and crucial for the birth of our universe.  The necessity of the multiverse for understanding the birth of our universe,
 is based on the arguments presented in \cite{laurarich, lauramulti} and sketched in Sec.2.A. Independently of time's enigma, the need for extending physics to the multiverse comes from basic questions such as: how did our universe come into being with such a special initial state. Such questions can not be meaningfully asked without the framework of the multiverse \cite{lauramulti,laurarich}. Besides, the entanglement of our universe with its bath may have already proven its relevance by leaving testable imprints on astrophysical observations (see \cite{rudnick,kashlinsky} and footnote $1$).

The existence of fundamental time in the multiverse becomes a logical consequence \cite{rudy} when taking the view that the underlying reality, the bath in which our universe is a small domain,  is relevant to our study of fundamental questions about nature. In fact, the opposite view that multiversal time may not exist, and the implication that the underlying reality of the multiverse is irrelevant, could lead to a 'Loschmidt'- type paradox and obscure our understanding of entropy, time and arrow's emergence. Similar to the situation arising from the 'molecular chaos' assumption, if multiversal time does not exist then local observers infer that the universe is a closed system with self-contained correlations. Such an assumption then leads to an information loss 'sneaked in' by construction - by ignoring the information 'hidden' in correlations between the multiverse and the universe, and in the gravitational sector - thereby creating an artificial, instead of a physical, asymmetry between initial and boundary conditions \cite{price}.

The view that the {\it multiverse is a closed system} but the {\it universe is an open system} entangled with the multiverse bath naturally leads to the second option, namely:   fundamental time exists. Then, conservation of information in the multiverse results in the reversability symmetry of fundamental time. Which implies, time in the multiverse is arrowless. Universally laws of physics carry  this time-reversal symmetry. Energy conservation would imply time-translation symmetry. This option leads us to conclude that multiversal time is fundamental, it has no direction, no beginning and no end.

{\it Local time at the position of the nucleating universe, although related, is not the same as the fundamental time of its underlying bath}. Entanglement with the multiverse and the coupling between the matter and gravitational D.o.F's, mentioned in Sec.2.A. and derived in \cite{laurarich,tomo}, drives a dynamical evolution of correlations. That is, the universe is an open and out-of-equlibrium system. Initially, the wavepacket has a superposition of geometries. As the bath 'pins down' the branches it entangles  with, (the system decohering), then there is a flow of information not only between the matter and gravitational sectors but also to the multiverse. This information is contained in the off-diagonal terms of the reduced density matrix for our branch of the wavefunction that describes how fast the superposition of different gemoetries decohere from each-other, as a result of entanglement with the bath \cite{laurarich}. Other channels of information loss are given by the intrinsic interaction of matter with gravitational D.o.F.s, such as, particle creation from curved spacetime, which describes a transfer of information from the varying gravitational fields with zero entropy to the matter sector, as well as the generic coupling of matter to curvature, (gravity), contained in Einstein equations. These channels contain the excitations of the gravitational vacuum correlated nonlinearly to matter. Despite some intriguing attempts \cite{grav}, the issue of gravitational entropy and its information transfer to the particle sector is still elusive and will be considered in a subsequent paper. From the local observers point of view, more  and more correlations 'hide' as irrelevant when the bubble goes through the nonequlibrium dynamics of expansion and decoherence. The  information is lost to the bath and the gravitational sector. As the universe grows, local observers in the branch  continue to lose information about the underlying reality, which breaks the reversal symmetry of time locally - the 'hidden' information is contained in the entanglement, information about the fact that this bubble is part of a bigger phase space, the multiverse.  From the bubble's perpective, the information is lost in a non-reversible way due to the local nonequlibrium dynamics and decoherence for reasons described next. Such dynamics guarantees that the untangling of our branch from the multiverse bath does not occur, thus the irreversability of the process. 

Time's arrow emerges only locally because time reversal symmetry is broken locally only, at the bubble nucleation, although the fundamental time of the multiverse is arrowless. If local observers were able to move away from the trajectory of the entangled and decohered branch, they would find time-symmetry restored away from the bubble. Thus, local breaking of time-reversal symmetry is due to the correlation changes between the system and the bath, which is driven by the gravitational dynamics of the system. Such local nonequilibrium, irreversable dynamics induces an asymmetry between the starting point, the initial state of the universe on the multiverse and the final state of the system. The system undergoes nonequlibrium dynamical evolution, which renders {\it its phase space nonergodic}, and ensures that the system can {\it never return to its initial state} \cite{laurarich}. We perceive this change in the correlations of the system from the bath as a separation of the system from the multiverse and deduce locality since the observer in the system defines the relevant degrees of freedom locally.

In the birth of the universe from the multiverse scenario \cite{laurarich,lauramulti,tomo}, the initial conditions are not 'hand-picked'. Rather, they are dynamically  superselected from a generic set. The high energy, out-of equlibrium, initial system then tries to drive towards a symmetric final state, thereby driving an increase in entropy. The superselection rule for the initial states derived in \cite{laurarich}, separates 'terminal' from 'survivor' universes, by wiping out the former from the phase space available to our initial states. As a consequence of the superselection arising from the nonequilibrium dynamics and entanglement to the multiverse, phase space is not ergodic and Poincare recurrences do not occur. The implication is that the system can never return to its initial state. Thus the symmetry between the initial and final state can not be restored, resulting in an emerging arrow in the bubble. The asymmetry between the initial and boundary conditions, in this theory \cite{laurarich}, is not an artifact of breaking the symmetry by placing arbitrary conditions on the initial state while ignoring the boundary conditions, as rightly critized by H. Price \cite{price}. The asymmetry is governed and driven by the superselection rule on the multiverse arising from nonequlibrium dynamics of matter and gravitational D.o.F.'s.
This reasoning remains valid for contracting universes thus a reversal of time's arrow during the transition from an expanding to a contracting phase in an open system, such as the universe, also can not occur.

\subsection{Time-Reversal Symmetry for the Laws of Physics}

 Although fundamental time in the multiverse from which the universe nucleated, has no arrow, the local observer experiences that an arrow of time has emerged at the bubble due to the information of the entanglement with the bath lost and
hidden to the gravitational sector. At the bubble, time reversal symmetry is broken by the very act of nucleation and entanglement with the bath, since there is information loss about the underlying reality of the multiverse and about the gravitational entropy. As the bubble decoheres such entanglement with the bath is deemed as irrelevant, and these correlation ignored. The initial state is selected dynamically by the underlying physical laws. The nonequlibrium dynamics also ensures the nonergodicity of phase space which induces an asymmetry between the initial and boundary conditions - the bubble can never recur to its initial state.

However, the system inherits the same laws of physics from the multiverse that were valid before its decohering, without processing or changing them. But since globally fundamental time has no arrow, and the laws of physics therefore are symmetric with respect to time reversal operations, then each nucleating universe inheriting these laws from birth to a multiverse, would carry the same time-reversal symmetry for their laws. The time reversal symmetry of laws is a direct consequence of the fundamental time in the multiverse and not the local time in the bubble. For this reason, the emerging arrow of time at the bubble location and the time-symmetry of the laws of physics are concomitant since they are independent in origin. Unlike the arrow of time, laws do not emerge at the bubble nucleation, they are inherited from the underlying theory. This addresses our second mystery in a cohesive way: the physical laws a universe is born with, can be time-symmetric despite the breaking of this symmetry locally that induces the emergence of the local time's arrow. The time-symmetry of laws inherited by 'survivor' universes does not imply that physical laws are the same in every bubble. It simply makes a statement about a feature they all have in common from their origin in the multiverse, they share the time-symmetry property. 

Here we have demonstrated how, within a multiverse framework, we can achieve a coherent existence of both phenomena, an emerging local arrow of time and time symmetric laws. The separation of the system from the bath produces an arrow of time but does not modify or process the physical laws.

\section{Discussion } 
The  situation with time's enigma and the reversal symmetry of physical laws  is similar to the resolution of the Loschmidt paradox concerning Boltzmann's H-theorem. The reason for the entropy increases in Boltzmann's approach was the assumption of 'molecular chaos', which ignored the correlations and information about interaction and the microdynamics of particles. A similar situation arises in our case. Based on classical results of \cite{hawking} that assume equlibrium for the closed system, the gravitational entropy is usually taken to be zero, except for objects with horizons, such as DeSitter geometries and black holes. Yet the entropy of particles created from these gravitational fields in the universe is not zero. Besides, the interaction between the matter and gravitational sector is always present, which ensures that the open system remains out-of-equlibrium. The transfer of information from the particle sector to the gravitational sector, for the open system immersed in a bath, results in a loss since the role and the nonequilibrium dynamics of gravitational degrees of freedom is not taken into account. The assumption of independence of the system from the multiverse bath, together with information transferred to the gravitational sector and contained in the gravitational entropy are ignored as irrelevant. Information lost via these channels by enforcing locality, equilibrium, and choosing local matter D.o.F.'s as the only relevant D.o.F's, breaks time-reveral symmetry and leads to the emergence of a local arrow of time. In \cite{laurarich,lauramulti,tomo} we showed how to incorporate the superhorizon nonlocal entanglement with the bath, into the system's correlation. A general approach to quantifying correlations and information loss to the gravitational sector would require an understanding of the role of gravitational D.o.F's to Boltzmann's kinetic equation and H-function. Once we have a handle on the information lost via correlations of the particle with the gravitational sectors, we would be in a position to test the theory. The coupling between the matter and gravitational sectors results in nonequilibrium dynamics with the rate of information transferred from one sector to the other providing a concept of clocks. Clocks could not be built in a universe in perfect equlibrium, such as the thermal bath of radiation of a pure DeSitter geometry since in this case thermal equilibrium requires that entropy remains a constant, and that particle creation from the gravitational vacuum excitations becomes extinct.

A universe nucleating from the multiverse through the dynamic selection of its initial conditions results in the following picture: time in the multiverse is arrowless, with no beginning, no end and it is a fundamental building block. Laws of physics inherit the time-symmetry of the underlying theory. But local time in the bubble universe is a dynamic parameter with an emerging arrow at birth, since more and more information is lost about the underlying reality and transferred to the gravitational sector, which creates an asymmetry between the initial and final states of the local universe. The bubble inherits the arrowless physical laws despite it breaking the time-symmetry. The initial conditions of the bubble are dynamically chosen from physical laws from a generic state. The bubble tries to drive towards symmetric boundary conditions. Therefore the emerging arrow is not a consequence of an artificially imposed symmetry breaking between the selection of a preferred initial condition, without the selection of boundary conditions \cite
{price}. Initial and boundary conditions are both governed and dynamically selected by physical laws from generic sets, which can be achieved when the birth of the universe is studied within a multiverse framework with a fundamental arrowless time. The asymmetry arises from the information loss to the universe and the nonequlibrium dynamics of matter and gravitational D.o.F's that leads to a nonergodic phase space. Such asymmetry renders local time to be dynamic and have an emerging arrow.

\begin{figure}[!htbp]
\begin{center}
\raggedleft \centerline{\epsfxsize=4.1in \epsfbox{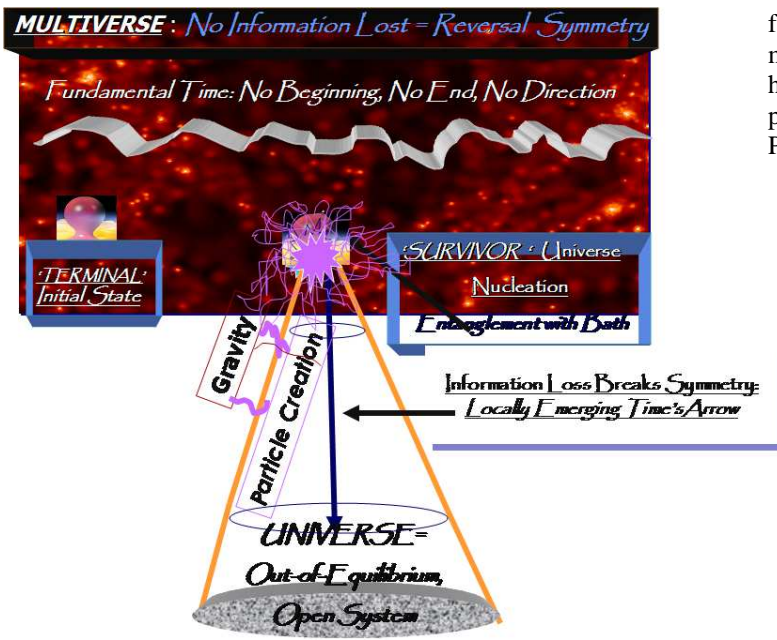}}
\caption{A schematic drawing of the birth of the universe from the multiverse. Only the high energy initial states in the multiverses are dynamically selected to give birth to a universe. Low energy states become 'terminal'. Information is conserved in the multiverse thus multiversal time is fundamental and directionless. The nucleating universe is an open system. Its out-of-equlibrium dynamics and entanglement with the multiverse break the time-reversal symmetry locally. But physical laws of the bubble originate from the multiverse thus carry the reversal-symmetry.}
\label{fig:fig2}
\end{center}
\end{figure}

As I tried to caution at the start, any attempts at tackling time's enigma remain in the realm of speculation until the calculational tools of information transfer and gravitational entropy are discovered. Without these tools it is hard to make testable predictions of the theory since the information contained in the interactions between matter and gravitational vacuum can not be estimated. Yet, the theory described in this letter, for time's enigma in the context of the dynamically selected birth of the universe from the multiverse, provides a coherent picture of the concomitant co-existence of the three aspects of this enigma: a locally emerging time's arrow, from a fundamentally arrowless time, for a universe that inherits the time-symmetry of its laws from the multiverse.

%begin{equation}
%tau = \frac{1}{4\pi A M_{0}\rho_{\infty}(1+w)} \ ,
%label{tau}
%end{equation}

%begin{eqnarray}
%& & R^3 \frac{dR}{dt} = {-64 \over 5} \frac{2 m_{0}^3}{ [1- 2 L t m_0]^3 }  \nonumber \\
%&- & \left[\frac{- 4 L  R^4 m_{0} }{[1 - 2 L t m_{0}] }  + 8 L R^6  \right ] \ ,
%label{diffeqn}
%end{eqnarray}

\medskip

Acknowledgment:  L.M.H is grateful to A.Guth for useful discussions and R.Vaas for valuable comments on the manuscript. L.M.H would like to thank DAMTP for their hospitality during the time this work was done. L.M.H  is supported in part by DOE grant DE-FG02-06ER1418, NSF grant PHY-0553312 and an FQXI grant.      .

\pagebreak

\end{document}